\documentclass{article}
\usepackage[latin9]{inputenc}
\usepackage{geometry}
\usepackage{float}
\usepackage{booktabs}
\usepackage{amsmath}
\usepackage{amssymb}
\usepackage[unicode=true,pdfusetitle,
 bookmarks=true,bookmarksnumbered=false,bookmarksopen=false,
 breaklinks=false,pdfborder={0 0 1},backref=section,colorlinks=false]
 {hyperref}
\usepackage{breakurl}

\makeatletter

\providecommand{\tabularnewline}{\\}

\usepackage{graphicx}
\usepackage{grffile}
\usepackage{siunitx}
\usepackage{multirow}
\usepackage{footnote}
\usepackage{breakurl}

\providecommand{\tabularnewline}{\\}

\usepackage{fullpage}\usepackage[justification=centering]{caption}\usepackage[all]{hypcap}\usepackage{microtype}% % % % % % % % % % % % % % % % % % % % % BEGIN DOCUMENT

\makeatother

\begin{document}

\title{Consequences of omitting spin-orbit partner configurations on B(E2)'s
and quadrupole moments in nuclei}

\author{L. Zamick$^{1}$, Y.Y. Sharon$^{1}$, S.J.Q. Robinson$^{2}$, M. Harper$^{1}$\\
 \textit{$^{1}$Department of Physics and Astronomy, Rutgers University,
Piscataway, New Jersey 08854, USA}\\
 \textit{$^{2}$Department of Physics, Millsaps College Jackson MS
39210} }

\date{\today}

\maketitle
 
\begin{abstract}
Calculations of B(E2)'s and quadrupole moments in the shell $g_{9/2}$
region below $^{100}$Sn are hampered by the fact that the inclusion
of the $g_{7/2}$ configuration leads to model spaces that are too
large to handle. Understanding the impact of specific orbit space functions in large-scale shell-model (LSSM) calculations helps to shed light on the validity of the results that are obtained. We therefore examine lighter nuclei if the $f-p$ region
where one can easily include all the orbitals, $f_{7/2}$, $p_{3/2}$,
$p_{1/2}$ and $f_{5/2}$. We perform such calculations but then take
a step back and exclude the $f_{5/2}$ orbital. By comparing the two
calculations we can hope to get insight into the importance of the
missing spin-orbital partner in other regions. 
\end{abstract}

\section{Introduction and Motivation}

We have recently performed calculations of B(E2)'s, static quadrupole
moments and other properties in both the $f-p$ shell region and the
$g_{9/2}$ shell region below $A$=100 \cite{robinson14}. In the latter
case we were unable to include the spin-orbital partner of $g_{9/2}$,
$g_{7/2}$. To study the implications of omitting a spin-orbit partner
we therefore retreat to the $f-p$ shell. Here we perform calculations
of B(E2)'s and static quadrupole moments involving the lowest $0^{+}$,
$2^{+}$, and $4^{+}$ states of selected even-even nuclei. We perform
the calculations with and without the inclusion of the $f_{5/2}$
orbit and compare the results.

\section{Shell Model Spaces and Effective Interactions}

The present calculations were performed with the Antoine shell model code\cite{antoine}. The full calculations were carried out in the shell model-space consisting of the $f_{7/2}$, $p_{3/2}$, $p_{1/2}$
and $f_{5/2}$ orbitals for both protons and neutrons. Thus, we included both spin-orbit partners for both the $p$ and $f$ orbitals.

The results are presented in Tables \ref{table:fpd6} and \ref{table:gx1}.
We give ratios with and without the $f_{5/2}$ orbit removed. 

It should be pointed out that we previously performed calculations
in Ge isotopes \cite{robinson11}. These calculations used a space consisting of the $f_{5/2}$, $p_{3/2}$, $p_{1/2}$, and $g_{7/2}$ orbitals for both protons and neutrons. The spin-orbit partner orbit $g_{7/2}$ was excluded and the spin-orbit partner orbital $f_{7/2}$ was part of a non-participating inert core. We were able to compare our calculated results with the experimental data for the Ge isotopes. Whereas overall the B(E2)'s were in good agreement with experiment the quadrupole moments were not.
In some cases the calculated and experimental Q's had opposite signs.
This is one of the main motivations for the present study, as we attempt to see whether the above discrepancy was due to the omission of the spin-orbit partner orbitals $g_{7/2}$ and also $f_{7/2}$. The results of the calculation depend on the effective interaction that is used. We employed two such interactions commonly used for $fp$ shell nucleus--FPD6\cite{fpd6} and GXPF1\cite{gx1}. Comparing the results that are obtained with the two interactions with each other and with the measured values acts as a check, both on the quality of each interaction and on the appropriateness of the shell model space that is used. In the Ge calculations \cite{robinson11} it was thought that the specific truncated shell-model space contributed more than the interaction details to the discrepancies with the experimental data. In \cite{robinson11} the large-scale shell-model wave-functions were very fractionated indicating the need to take collective effects and effective charges into account. Accordingly, in the present paper, we consider the static quadrupole moments, Q, and the reduced electric transition probabilities B(E2)'s. These properties provide information about the collectivity of nuclei. The Q of a state tell us if the nucleus in that state is oblate or prolate. The B(E2)'s are often known to better precision than the Q's and they have been measured more frequently. Large B(E2) values indicate collectivity but they cannot tell us about prolateness or oblateness.

\section{Results and Discussion}

In Table 1 we present the calculated values of the quadrupole moments Q($2_1^+$) and Q($4_1^+$) in $e(fm)^2$ and the reduced transition probabilities B(E$2;2_1^+\rightarrow0_1^+$) and B(E$2;4_1^+\rightarrow2_1^+$) in $e^2(fm)^4$. The results are for the FPD6 and GXPF1 interactions and for the five nuclei: $^{44}$Ti, $^{46}$Ti, $^{48}$Ti, $^{48}$Cr, and $^{50}$Cr. Each result is calculated twice, once in the full $fp$ space and once in the truncated $fp$ space, with the $f_{5/2}$ orbital missing. The effective charges that were used in these calculations were $e_p=1.33$ and $e_n=0.64$ in \cite{fpd6} and $e_p=1.5$ and $e_n=0.5$ in \cite{gx1}.

\begin{savenotes}
\begin{table} [ht]
\centering \protect\protect\caption{Calculated values for the quadrupole moments and B(E2) values in the
full $fp$ model space and in the truncated space which is missing the
$f_{5/2}$ orbit. Results are listed for both FPD6/GXPF1. The units
are $e(fm)^2$ and $e^2(fm)^4$ respectively.}
\hspace{-2.cm}
\small\setlength{\tabcolsep}{2pt}
\begin{tabular}{llllllr}
\toprule 
\addlinespace
\protect Ratio  & $^{44}$Ti  & $^{46}$Ti  & $^{48}$Ti  & $^{48}$Cr  & $^{50}$Cr  & $^{52}$Fe \tabularnewline
\addlinespace
Q$(2_{1}^{+})_{\text{full}}$  & -21.572 /-5.133  & -23.505 /-12.720  & -18.807 /-13.719  & -35.416 /-30.160  & -32.914 /-28.34  & -33.642 /-30.451 \tabularnewline
Q$(2_{1}^{+})_{\text{trunc}}$  & -15.161/ 5.147  & -20.756 /-7.59  & -16.532 /-10.904  & -30.046 /-24.6379  & -28.025 /-24.746  & -26.906/-24.674 \tabularnewline
Q$(4_{1}^{+})_{\text{full}}$  & -28.918 /-16.378  & -31.02 /-22.985  & -20.724 /-11.425  & -45.47 /-40.425  & -41.867 /-35.74  & -38.487 /-37.564 \tabularnewline
Q$(4_{1}^{+})_{\text{trunc}}$  & -19.157 /-3.068  & -26.323 /-13.626  & -15.894 /-7.9999  & -38.541 /-31.872  & -34.754 /-29.343  & -32.04 /-29.348 \tabularnewline
B(E2;$2\rightarrow0)_{\text{full}}$  & 139.12 / 102.56  & 155.321 / 122.137  & 127.04 / 103.446  & 312.37 / 244.01  & 274.96 / 211.82  & 291.097 / 218.04 \tabularnewline
B(E2;$2\rightarrow0)_{\text{trunc}}$  & 119.629 / 84.50  & 130.122 / 104.26  & 104.55 / 91.8394  & 221.932 / 179.22  & 196.488 / 163.866  & 171.83 / 141.64 \tabularnewline
B(E2;$4\rightarrow2)_{\text{full}}$  & 189.47 / 132.46  & 220.395 / 154.380  & 192.95/ 146.320  & 436.24/ 329.60  & 398.93 / 301.16  & 425.124 / 284.414 \tabularnewline
B(E2;$4\rightarrow2)_{\text{trunc}}$  & 156.688 / 109.38  & 182.444 / 129.23  & 158.833 / 126.0289  & 311.330 / 250.97  & 277.164 / 227.2459  & 226.489 / 180.13 \tabularnewline
\end{tabular}\label{table:all} 
\end{table}
\end{savenotes}
We note that in Table 1, with the exception of Q$(2_1^+)_{\text{trunc}}$ for $^{44}$Ti, all the quadrupole moments are negative. These results indicate prolate intrinsic shapes for all the five nuclei which lie in the beginning of the the $fp$ shell. Thus, overall this specific aspect of the result is not affected by the truncation. We also observe that for all the nuclear properties in Table 1, Q's and B(E2)'s alike, the magnitude of the calculated FPD6 result is larger that the corresponding GXPF1 results in both the full and truncated spaces.

In Table 2 (for the FPD6 interaction) and in Table 3 (for the GXPF1 interaction) we compare the results in the full and truncated spaces. We consider the ratio of squares of the corresponding quadrupole moments in each of the two spaces as well as as the ratio of the corresponding B(E2) values.

\begin{savenotes} 
\begin{table}[H]
\centering \protect\protect\caption{Quadrupole and B(E2) ratios using the FPD6 interaction.}

\begin{tabular}{rrrrrrr}
\toprule 
\addlinespace
\protect Ratio  & $^{44}$Ti  & $^{46}$Ti  & $^{48}$Ti  & $^{48}$Cr  & $^{50}$Cr  & $^{52}$Fe \tabularnewline
\addlinespace
$\left(\dfrac{\text{Q}(2_{1}^{+})_{\text{full}}}{\text{Q}(2_{1}^{+})_{\text{trunc}}}\right)^{2}$  & 2.02  & 1.28  & 1.29  & 1.39  & 1.38  & 1.56 \tabularnewline
$\left(\dfrac{\text{Q}(4_{1}^{+})_{\text{full}}}{\text{Q}(4_{1}^{+})_{\text{trunc}}}\right)^{2}$  & 2.28  & 1.39  & 1.70  & 1.39  & 1.45  & 1.44 \tabularnewline
$\dfrac{\text{B}(\text{E2};2\rightarrow0)_{\text{full}}}{\text{B}(\text{E2};2\rightarrow0)_{\text{trunc}}}$  & 1.16  & 1.19  & 1.21  & 1.41  & 1.40  & 1.77 \tabularnewline
$\dfrac{\text{B}(\text{E2};4\rightarrow2)_{\text{full}}}{\text{B}(\text{E2};4\rightarrow2)_{\text{trunc}}}$  & 1.21  & 1.22  & 1.21  & 1.40  & 1.44  & 1.87 \tabularnewline
\addlinespace
 &  &  &  &  &  & \tabularnewline
\end{tabular}\label{table:fpd6} 
\end{table}

\end{savenotes} \raggedright

\begin{savenotes} 
\begin{table}[H]
\centering \protect\protect\caption{Quadrupole and B(E2) ratios using the GXPF1 interaction.}

\begin{tabular}{rrrrrrr}
\toprule 
\addlinespace
\protect Ratio  & $^{44}$Ti  & $^{46}$Ti  & $^{48}$Ti  & $^{48}$Cr  & $^{50}$Cr  & $^{52}$Fe \tabularnewline
\addlinespace
$\left(\dfrac{\text{Q}(2_{1}^{+})_{\text{full}}}{\text{Q}(2_{1}^{+})_{\text{trunc}}}\right)^{2}$  & 0.996  & 2.81  & 1.58  & 1.50  & 1.31  & 1.52\tabularnewline
$\left(\dfrac{\text{Q}(4_{1}^{+})_{\text{full}}}{\text{Q}(4_{1}^{+})_{\text{trunc}}}\right)^{2}$  & 28.516  & 2.84  & 2.04  & 1.61  & 1.44  & 1.63 \tabularnewline
$\dfrac{\text{B}(\text{E2};2\rightarrow0)_{\text{full}}}{\text{B}(\text{E2};2\rightarrow0)_{\text{trunc}}}$  & 1.21  & 1.17  & 1.13  & 1.36  & 1.29  & 1.54 \tabularnewline
$\dfrac{\text{B}(\text{E2};4\rightarrow2)_{\text{full}}}{\text{B}(\text{E2};4\rightarrow2)_{\text{trunc}}}$  & 1.21  & 1.19  & 1.16  & 1.31  & 1.33  & 1.58 \tabularnewline
\addlinespace
 &  &  &  &  &  & \tabularnewline
\end{tabular}\label{table:gx1} 
\end{table}
\end{savenotes} \raggedright

It should be noted that all of the rations in Tables 2 and 3 are greater than one. This indicates that for all the nuclei considered, and for both interactions, eliminating the spin-orbit partner orbital from the shell model space results in smaller Q and B(E2) values and thus less collectivity.

Tables 2 and 3, but not Table 1, include calculated results for the nucleus $^{52}_{26}$Fe$_{26}$, with more protons than any other nucleus we considered. Typically just about all the ratio values are larger for this nucleus, than for any of the other nuclei considered here. This indicates that at least at the beginning of the shell, omitting the spin-orbit partner orbital, has a bigger impact as the number of protons increases.

The results of \cite{robinson14} for N=Z nuclei suggest that somewhat different results may be obtained in different orbitals ($g_{9/2}$ vs $f_{7/2}$ here). An important question is whether, for a given nucleus and interaction one can compensate for the omission of the spin-orbit partner orbital by simply rescaling, in a consistently compatible way the Q's and the B(E2)'s/ Tables 2 and 3 help to answer this question.

For a given nucleus and interaction we can define ``simple behavior'' as when the square of the ratio of the quadrupole moments (see Tables 2 and 3)is the same as the ratio of the corresponding B(E2)'s. In such a case one can say that the effect of excluding the $f_{5/2}$ orbital simply corresponds to a renormalization by an overall ``effective charge'' of the Q's and the B(E2)'s and thus, is not a serious matter

Let us illustrate with the results from Table 2 for $^{48}$Cr with the FPD6 interaction. All four of the ratios are essentially the same, at about 1.40, illustrating a very simple behavior. More specifically in this case B(E2)$_{\text{full}}$/B(E2)$_{\text{trunc}}=1.40$ and Q$_{\text{full}}$/Q$_{\text{trunc}}=\sqrt{1.40}=1.183$. 

Computationally, for a nucleus that behaves ``simply'' we can first calculate a Q$_{\text{trunc}}$ and a B(E2)$_{\text{trunc}}$ in the truncated space, a simpler calculation. Then we would calculate the Q$_{\text{full}}$ in the full space and find the value of the ratio of Q$_{\text{full}}$/Q$_{\text{trunc}}$. Finally, we would multiply the B(E2)$_{\text{trunc}}$ by the square of this ratio value to obtain the B(E2)$_{\text{full}}$ without doing the more complicated and computer time-consuming, direct calculation of the B(E2)$_{\text{full}}$.

According to Table 3 with FPD6, for the Q($2_1^+$) and the B(E2;$2_1^+\rightarrow0_1^+$), simple behavior is exhibited by$^{46}$Ti, $^{48}$Ti,$^{48}$Cr, and $^{50}$Cr within $8\%$ or less, and within $13\%$ by $^{52}$Fe, but not by $^{44}$Ti (with only two valence protons and two valence neutrons). From Table 3 we see that with GXPF1 such simple behavior is exhibited with $10\%$ by the heavier nuclei: $^{48}$Cr, $^{50}$Cr, and $^{52}$Fe. Indeed, for $^{44}$Ti the small Q($2_1^+$)'s have different signs for the truncated and full calculation. Small values of Q may indicate the lack of collectivities or else vibrational behavior.

It is interesting to note, from Tables 2 and 3, that for both interaction and for every nucleus considered in these tables, the ratios for the B(E2;$4_1^+\rightarrow2_1^+$) and for the B(E2;2$_11^+\rightarrow0_1^+$), (the last two lines of each table) agree to within at least $6\%$. The Q($4_1^+$) and Q($2_1^+$) ratios often agree to within $12\%$, except for $^{44}$Ti with the GXPF1 interaction and $^{48}$Ti with both interactions.

We have shown that for all the nuclei under consideration eliminating the spin-orbit partner orbital $f_{5/2}$ results in smaller Q's and B(E2)'s and thus, in less collectivity. If all nuclei exhibited very simple behavior, like the ``gold standard'' nucleus $^{48}$Cr with the FPD6 interaction, we could simulate the effect of the missing orbital by simply adjusting the overall effective charge. This usually seems to be the case for the B(E2)'s. For the Q's it is sometimes, but not always true, so one needs to be more careful there. 

 It is of interest to relate the present work to the well cited  works of P. Federman and S. Pittel \cite{Pittel 77,Pittel 79}. In their shell model studies of nuclear deformation in the Zr and Mo isotopes they conclude that deformations become large when the $T=0$ neutron-proton interaction dominates over the $J=0$ $T=1$ pairing interaction. They further emphasize that for light and medium mass nuclei the simultaneous occupation of spin-orbit partners plays a crucial role in determining the onset of nuclear deformation. In agreement with their works we certainly get much stronger deformation when we allow occupation of spin-orbit partners, as manifested in increased B(E2)'s. However, we ask whether the exclusion of one of the spin-obit partners can be simulated by a simple increase of the effective charge. We are motivated by the practical consideration that in heavier nuclei the shell model spaces become prohibitively large when one attempts to include both members. Our results are somewhat mixed but we get the encouraging result e.g. in $^{48}$Cr that when the B(E2)'s are large this method works fairly well. We also consider static quadruple moments which we feel are more sensitive to details. For example, in both the rotational model and vibration model the B(E2) from say $2^+$ to $0^+$ is large, but the static quadruple moment, though large in the rotational model is zero in the simplest version of the rotational model.

S.J.Q.R. thanks Millsaps College for a Faculty Development Grant award. Y.Y.S. acknowledges a Research and Professional Development Grant from the Richard Stockton College of New Jersey. Matthew Harper thanks the Rutgers Aresty Research Center for Undergraduates for support during the 2014-2015 fall-spring session.
%\end{section}

\end{document}